\documentclass{myaa}

\usepackage[varg]{txfonts}

\usepackage{natbib}
\usepackage{graphicx}
\usepackage{epstopdf}
\usepackage{amssymb}
\usepackage{float}

\begin{document}

\title{Conundrums and constraints concerning the formation of distinct families of small bodies in our solar system -- An alternative view}

\author{Dimitris M. Christodoulou\inst{1,2}  
\and 
Demosthenes Kazanas\inst{3}
}

\institute{
Lowell Center for Space Science and Technology, University of Massachusetts Lowell, Lowell, MA, 01854, USA.\\
\and
Dept. of Mathematical Sciences, Univ. of Massachusetts Lowell, 
Lowell, MA, 01854, USA. \\ E-mail: dimitris\_christodoulou@uml.edu\\
\and
NASA/GSFC, Laboratory for High-Energy Astrophysics, Code 663, Greenbelt, MD 20771, USA. \\ E-mail: demos.kazanas@nasa.gov \\
}


\def\gsim{\mathrel{\raise.5ex\hbox{$>$}\mkern-14mu
                \lower0.6ex\hbox{$\sim$}}}

\def\lsim{\mathrel{\raise.3ex\hbox{$<$}\mkern-14mu
               \lower0.6ex\hbox{$\sim$}}}

\abstract{We have proposed an alternative model for the formation of our solar system that does not predict any mean-motion resonant interactions, planetary migrations, or self-gravitating instabilities in the very early isothermal solar nebula and before the protosun has formed. Within this context of nonviolent protoplanetary evolution over more than 10 million years, we examine some conundrums and constraints that have been discovered from studies of small bodies in the present-day solar system (Jupiter and Neptune's Trojans and their differences from Kuiper belt objects, the irregular satellites of gaseous giants, the stability of the main asteroid belt, and the Late Heavy Bombardment). These issues that have caused substantial difficulties to models of violent formation do not appear to be problematic for the alternative model, and the reason is the complete lack of violent events during the evolution of protoplanets.}

\authorrunning{ }
\titlerunning{Conundrums and constraints about the formation of the solar system}

\maketitle


\section{Introduction}\label{intro}

In \cite{chr19}, we have advocated for an alternative model of the formation of our solar system, a model that does not support planet migrations, or self-gravitational instabilities, or any other form of violent evolution. Planets form in-situ and in relative safety inside multiple local gravitational potential wells provided by the isothermal solar nebula, long before the protosun is formed or the gaseous self-gravitating disk is gone. The model solar nebula is stable to self-gravity-induced instabilities as it rotates slowly and the radial excursions of the gaseous giants inside their potential minima are no more than a few AU. As a result of the particular nebular density and rotation profiles, the emerging planetary configuration exhibits no mean-motion resonances (MMRs) at any time, so after the gas has cleared and the Sun has formed, there are no instabilities to affect such a pristine and well-organized structure, pretty much the one we observe nowadays. The only evolution left to take place is a gradual adjustment to precise Keplerian orbits that is driven by the orbital angular momentum of each planet and that is not expected to displace the planets by much, since the mass interior to each orbit is approximately constant during the concluding stage of gas accretion on to the protosun.

Recent ALMA and SPHERE observations strongly support this picture of nonviolent formation \citep{alm15,and16,rua17,lee17,lee18,mac18,ave18,cla18,kep18,guz18,ise18,zha18,dul18,fav18,har18,hua18,per18,kud18,lon18,pin18,vdm19}. Virtually all gaseous disks observed show regular disk structures and the best-resolved disks show well-organized dark annular gaps, presumably carved out by protoplanets already circling around inside the dark gaps. These dark gaps are not arranged in mean-motion resonances, thus our prediction is that these young protoplanetary systems are dynamically stable and they will continue to be in this state for billions of years to come. In this sense, these well-organized systems are similar to our much older solar system. 

Before such a model of nonviolent planet formation opens the door to a new theory of formation, we need to return to our solar system and examine how various conundrums and constraints relating to the evolution of small bodies fit in such a novel context. We do so in \S~\ref{issues}, where we revisit the following aspects of formation of small bodies during the early evolution of our solar system: the Trojan asteroids of Jupiter and Neptune (\S~\ref{Trojans}); the irregular satellites of the gaseous giants (\S~\ref{irr}); the stability of the main asteroid belt (\S~\ref{belt}); and the hypothetical event termed ``Late Heavy Bombardment'' (LHB) (\S~\ref{lhb}). In \S~\ref{disc}, we summarize our conclusions.

\section{Conundrums and Constraints--Alternative Nonviolent Views}\label{issues}

Below we provide alternative nonviolent views of several aspects of the formation and evolution of small bodies in our solar system, views that do not at all rely on the Nice model \citep{tsi05,mor05,gom05} and its recent variant called ``gravitational instability'' \citep{nes18}; although these models have been successful in explaining some of the issues highlighted below, including the LHB, an event that most likely has never occurred in the history of our solar system \citep[e.g.,][]{zel17,nes18,cle18}; thus, it invalidates all previous models that managed to ``explain'' it. We stress that the ability of past models to explain the LHB casts doubt to all other numerical simulations preformed in the past from dubious or hypothetical initial conditions; it certainly appears that numerical modelers can rediscover any hypothetical features thrown at them, including ficticious events that have never occurred. This is not solid science, so trust in past numerical simulations must have been lost at the present time.

\subsection{The Trojans Conundrum}\label{Trojans}

The Trojan asteroids of Jupiter have been explained as captured objects by numerical simulations in the context of the Nice model and planetary migrations crossing low-order MMRs \citep{mor05,rob06}. Recently \cite{jew18a,jew18b} studied the Trojans of Neptune and found that the blue colors of the two populations are identical, suggesting a common origin. Unfortunately, the Kuiper belt cannot be the reservoir from which these objects originated because it is dominated by objects of very red color. This presents us with a conundrum: two gaseous giants with identical blue Trojan populations and no common source of origin. Surface modifications cannot be the explanation either because Neptune's Trojans exist at much colder temperatures that preclude such chemical evolution. 

\subsubsection{Trojan Asteroids--Alternative View}

In the context of our model, all Trojans are primordial material left over since the formation of the protoplanets. Once the giant cores are formed inside their potential wells, the Lagrangian L4/L5 points along each orbit are well-defined, and the protoplanets could not at all accrete this material that was constantly orbiting at $\pm$60 degrees away from them \citep[see also][]{mar02}. Excitation of inclinations and eccentricities took place during subsequent evolution of the system orbiting inside the same gravitational potential minimum and interacting through tidal forces as the protoplanet was growing. The fact that the Trojans are mostly blue in color gives us insight into the planetesimals orbiting inside the potential wells during the isothermal phase of the outer solar nebula. And the color similarity tells us that the solid cores of the gaseous giants had initially the same composition, which has been suspected in the past \citep[][and references therein]{ben09,for10,mil16}.

It has been argued that the Trojan populations and the Kuiper belt objects share roughly similar size distributions. As \cite{jew18a} describes, this is not a convincing result and we cannot trust it and assign a common origin to these objects. We add to it that, statistics aside, the raw data do not support populations of similar sizes: the Kuiper belt hosts nine objects that are larger and most more massive than the largest asteroid Ceres. Statistics cannot change our perception that the Kuiper belt is occupied by many minor planets, whereas the asteroid belt is definitely not.

\subsection{The Irregular Satellites Conundrum}\label{irr}

This conundrum was laid out long ago \citep{jew07,nes07}. The gaseous giants have more satellites in irregular orbits than in regular, nearly coplanar orbits. Furthermore, most of these irregular satellites are in retrograde orbits which would suggest that such orbits are more stable than the regular orbits of satellites formed in a primordial circumplanetary disk.

Once again, early capture of objects mostly in retrograde orbits is the favorite mechanism but this type of capture does not really work. These satellites are all bluer than Kuiper belt objects, so a common repository to provide the objects does not exist. Thus, there is no early part of the solar system that can supply so many blue objects and scatter them in ``convenient'' orbits, resulting in so many retrograde captures \citep[a total of 88 versus only 19 prograde captures;][]{jew07}.

\subsubsection{Irregular Satellites--Alternative View}

In the context of our model, most irregular satellites are primordial material left over since the formation of the protoplanets. But they were not formed in the accretion disks surrounding the protoplanets. They were formed inside the same gravitational potential wells as their associated protoplanets. Excitations of inclinations and eccentricities certainly took place as these objects interacted repeatedly with the major protoplanetary cores in two-body encounters. Furthermore, the blue colors of the objects testify to their common primordial nebular origin, just as in the case of the Trojans discussed above.

Objects orbiting interior to a major protoplanetary core may be captured in prograde orbits (as the major core overtakes them); on the other hand, objects orbiting at slightly larger heliocentric distances may be captured in retrograde orbits as they overtake the orbiting major core. It is known that the local potential wells provided by the solar nebula are asymmetric and more extended toward larger heliocentric distances \citep{chr17}, so there is a lot more material available to be captured in retrograde orbits. Perhaps this asymmetric structure of the wells could explain why most captured irregular satellites are in retrograde orbits. The asymmetry of the wells diminishes quite substantially in the current locations of the two outermost giant planets \citep{chr19}, and this property of the early solar nebula may explain why the vast majority of retrograde satellites is found only around Jupiter (49 of 88) and, to a lesser degree, around Saturn (27 of 88) \citep[][their Table 2]{jew07}. The irregular satellites could not possibly have originated in the Kuiper belt because then Uranus and Neptune would have captured many of them and scattered the rest of the objects all over the solar system. But quite the opposite is observed, which lends support to in-situ captures of these objects by the major protoplanetary cores.

The capture mechanism of irregular satellites is not in doubt in the nonviolent scenario that we advocate. These objects were trapped inside the same gravitational potential wells as the corresponding major protoplanetary cores. Multiple interactions must have resulted in multiple captures and in excitations of inclination and eccentricity of the captured fragments. The main point here is that the irregular satellites could not escape from each protoplanetary core because they were orbiting inside the same potential trough. This alleviates the problem of different accretion scenarios between the two inner and the two outer gaseous giants brought to our attention by \cite{jew05}.

\subsection{The Asteroid Belt Constraint}\label{belt}

In numerical models of planetary migrations inspired by the Nice/instability model, one persistent thorn is the survival of the asteroid belt, the Hildas, and the Trojans. In some simulations, the main asteroid belt gets excited to larger inclinations and eccentricities, but it survives in some reasonable form; but the Hildas and the Trojans do not survive \citep{mor09,min11,roi15}. This empirical fact alone should have had us thinking that violent evolution has not occurred in our solar nebula.

\subsubsection{Asteroid Belt--Alternative View}

In the context of our model, none of the above thorny issues materializes. In the absence of extensive migration, the Hildas and the Trojans may easily survive. The main asteroid belt may not have been able to form one major protoplanet during the early isothermal evolution of the nebula, but its objects were all confined safely within a potential trough. So if Jupiter's tidal forcing was responsible for continually stirring the asteroids, tidal forces would still have been unable to knock them out of their potential minimum and destroy the asteroid belt altogether. Furthermore, the temperature in the main asteroid belt was high enough for chemical processes to set in and differentiate the population over billions of years. This explains why the asteroid belt is made of both blue and red objects and some (D-type) asteroids have similar spectral properties to those of the Trojans. There is no need to invoke inward planetesimal migrations and contamination of the main belt \citep{lev09,vok16}, a scenario that already sounds far-fetched: in this scenario, the planetesimals travel inward for many AUs and then, for reasons unknown, they somehow circularize inside the main asteroid belt and contaminate its composition.

\subsection{The Late Heavy Bombardment is No Longer a Constraint}\label{lhb}

In our solar system, a spike in asteroids flying inward called the LHB has been hypothesized to have occurred $\sim$3.8 Gyr ago. This cataclysmic event was invented in order to explain craters on the moon's surface that appear to have the same age and it forms one of the cornerstones of the Nice model. This is because, in the absence of localized potential wells, extensive migration of the gaseous giants could hurl many asteroids inward and have them rain down on the objects of the inner solar system.

\subsubsection{Late Heavy Bombardment--Alternative View}

The LHB has recently been disputed both on observational grounds and by numerical simulations \citep{rey15,kai15,zel17,nes17,nes18,cle18}. Thus, another violent event has been invalidated. To be sure, once the gaseous disk was gone and the planets emerged in slightly non-Keplerian rotation, they had to adjust their orbits; and this event may have caused planetesimals and primordial asteroids to rain down to the inner planets and the moon. But there was no spike of activity 3.8 Gyr ago. Rather the event took place on long secular timescales as the gaseous giants were adjusting continually to the changing gravitational potential, just as the latest discoveries indicate \citep[e.g.,][]{zel17,cle18}.

\section{Summary}\label{disc}

In view of an alternative model of nonviolent planet formation in our solar nebula \citep{chr19}, we have revisited certain conundrums and constraints dictated by the smaller present-day objects in our solar system (\S~\ref{issues}). We considered:
\begin{itemize}
\item[1.]The blue color conundrum of the Trojan asteroids of Jupiter and Neptune.
\item[2.]The irregular, mostly retrograde, satellites of the gaseous giants.
\item[3.]The constraint posed by the long-term survival of the main asteroid belt, the Hildas, and the Trojans.
\item[4.]The Late Heavy Bombardment of the objects in the inner solar system.
\end{itemize}

It appears that it is not a challenge to explain these issues in a nonviolent scenario of well-organized, in-situ protoplanet formation confined inside multiple localized gravitational potential wells provided by the isothermal solar nebula itself during its very early isothermal evolution. We find that:
\begin{itemize}
\item[(a)]The blue colors of Trojans and irregular satellites can be explained if these objects were formed in-situ, inside multiple local gravitational potential wells that also hosted the major protoplanetary cores. The common repository of these objects was certainly not the Kuiper belt, it was the outer solar nebula itself. The same type of material also formed the solid cores of the gaseous giants.
\item[(b)] In a nonviolent planet formation scenario, there is no issue about the survival of the main asteroid belt inside its own potential well or about the LHB, an event that most likely did not ever take place. The Hildas and the Trojans also reside in the potential well in which Jupiter's core was formed, and there were no strong perturbations to cause ejections and scattering all over the early solar system.
\end{itemize}

It is long overdue for us to take another look at the present structure of our solar system; all along it has been telling us that extensive planet migrations and MMR-crossing interactions did not occur in the solar nebula. We now have strong confirmation of such nonviolent evolution occurring in many protoplanetary disks observed by ALMA and SPHERE (citations listed in \S~\ref{intro}), unlike in the fully-formed exoplanetary systems discovered in our solar neighborhood (a severely distance/volume limited sample), most of which have been disturbed or expunged by planet migrations and/or instabilities (\url{www.exoplanets.org}).

The fact that the ALMA disks show well-organized dark annular gaps (presumably carved out by orbiting protoplanets) also resolves another major issue: our solar system is not at all special in the grand scheme of things. There are many other young systems, currently evolving past their early isothermal phase, in which protoplanets do not suffer from MMR interactions or self-gravitational instabilities and continue to grow in relative safety for at least 1-10 millions of years.

\begin{thebibliography}{}

\bibitem[ALMA Partnership(2015)]{alm15}ALMA Partnership, Brogan, C. L., P\'erez, L. M., Hunter, T. R., et al. 2015, \apjl, 808, L3

\bibitem[Andrews et al.(2016)]{and16}Andrews, S. N., Wilner, D. J., Zhu, Z., et al. 2016, \apjl, 820, L40

\bibitem[Avenhaus et al.(2018)]{ave18}Avenhaus, H., Quanz, S. P., Garufi, A., et al. 2018, \apj, 863, 44

\bibitem[Benvenuto et al.(2009)]{ben09}Benvenuto, O. G., Fortier, A., \& Brunini, A. 2009, Icarus, 204, 752

\bibitem[Christodoulou \& Kazanas(2017)]{chr17}Christodoulou, D. M., \& Kazanas, D. 2017, Res. Astr. Astroph., 17, 129

\bibitem[Christodoulou \& Kazanas(2019)]{chr19}
Christodoulou, D. M., \& Kazanas, D. 2019, Part 1, arXiv: 1901.02593 (Solar Nebula)

\bibitem[Clarke et al.(2018)]{cla18}Clarke, C. J., Tazzari, M., Juhasz, A., et al. 2018, \apjl, 866, L6

\bibitem[Clement et al.(2018)]{cle18}Clement, M. S., Kaib, N. A., Raymond, S. N., \& Walsh, K. J. 2018, Icarus, 311, 340

\bibitem[\'Cuk \& Gladman(2006)]{cuk06}
\'Cuk, M., \& Gladman, B. J. 2006, Icarus, 183, 362

\bibitem[Dullemond et al.(2018)]{dul18}Dullemond, C. P., Birnstiel, T., Huang, J., et al. 2018, DSHARP VI, \apjl, 869, L46

\bibitem[Favre et al.(2018)]{fav18}Favre, C., Fedele, D., Maud, L., et al. 2018, \apj, arXiv:1812.04062

\bibitem[Fortney \& Nettelmann(2010)]{for10}Fortney, J. J., \& Nettelmann, N. 2010, Space Science Reviews, 152, 423

\bibitem[Gomez et al.(2005)]{gom05}Gomes, R., Levison, H. F., Tsiganis, K., \& Morbidelli, A. 2005, Nature, 435, 466

\bibitem[Guzm\'an et al.(2018)]{guz18}Guzm\'an, V. V., Huang, J., Andrews, S. M., et al. 2018, DSHARP VIII, \apjl, 869, L48

\bibitem[Harsono et al.(2018)]{har18}Harsono, D., Bjerkeli, P., van der Wiel, M. H. D., et al. 2018, Nature Astronomy, 2, 646

\bibitem[Huang et al.(2018)]{hua18}Huang, J., Andrews, S. N., Dullemond, C. P., et al. 2018, DSHARP II, \apjl, 869, L42

\bibitem[Isella et al.(2018)]{ise18}Isella, A., Huang, J., Andrews, S. N., et al. 2018, DSHARP IX, \apjl, 869, L49

\bibitem[Jewitt(2018a)]{jew18a}Jewitt, D. 2018a, DPS Meeting \#50, id. 200.06

\bibitem[Jewitt(2018b)]{jew18b}Jewitt, D. 2018b, ASPC, 513, 33

\bibitem[Jewitt \& Sheppard(2005)]{jew05}Jewitt, D., \& Sheppard, S. 2005, Space Sci. Rev., 116, 441

\bibitem[Jewitt \& Haghighipour(2007)]{jew07}
Jewitt, D., \& Haghighipour, N. 2007, ARA\&A, 45, 261

\bibitem[Kaib \& Chambers(2015)]{kai15}Kaib, N. A., \& Chambers, J. E. 2015, AAS DPS Meeting \#47, id 418.03

\bibitem[Keppler et al.(2018)]{kep18}Keppler, M., Benisty, M., M$\ddot{\rm u}$ller, A., et al. 2018, A\&A, 617, A44

\bibitem[Kudo et al.(2018)]{kud18}Kudo, T., Hashimoto, J., Muto, T., et al. 2018, \apj, 868, L5

\bibitem[Lee et al.(2017)]{lee17}Lee, C.-F., Li, Z.-Y., Ho, P. T. P., et al. 2017, \apj, 843, 27

\bibitem[Lee et al.(2018)]{lee18}Lee, C.-F., Li, Z.-Y., Hirano, N., et al. 2018, \apj, 863, 94 

\bibitem[Levison et al.(2009)]{lev09}Levison, H. F., Bottke, W. F., Gounelle, M., et al. 2009, Nature, 460, 364

\bibitem[Long et al.(2018)]{lon18}Long, F., Pinilla, P., Herczeg, G. J., et al. 2018, \apj, 869, 17

\bibitem[Mac\'ias et al.(2018)]{mac18}Mac\'ias, E., Espaillat, C. C., Ribas, \'A, Schwarz, K. R., et al. 2018, \apj, 865, 37

\bibitem[Marzari et al.(2002)]{mar02}Marzari, F., Scholl, H., Murray, C., \& Lagerkvist, C. 2002, Asteroids III, ed. Bottle, W. F. et al., Univ. of Arizona Press, 725

\bibitem[Militzer et al.(2016)]{mil16}Militzer, B., Soubiran, F., Wahl, S. M., \& Hubbard, W. 2016, J. Geoph. Res., 121, 1552

\bibitem[Minton \& Malhotra(2011)]{min11}
Minton, D. A., \& Malhotra, R. 2011, \apj, 732, 53

\bibitem[Morbidelli et al.(2009)]{mor09}
Morbidelli, A., Brasser, R., Tsiganis, K., Gomes, R., \& Levison, H. F. 2009, A\&A, 507, 1041

\bibitem[Morbidelli et al.(2005)]{mor05}
Morbidelli, A., Levison, H. F., Tsiganis, K., \& Gomes, R. 2005, Nature, 435, 462

\bibitem[Nesvorn\'y(2018) ]{nes18}Nesvorn\'y, D. 2018, ARA\&A, 56, 137

\bibitem[Nesvorn\'y et al.(2017) ]{nes17}Nesvorn\'y, D., Roig, F., \& Bottke, W. F. 2017, A\&A, 153, 103

\bibitem[Nesvorn\'y et al.(2007)]{nes07}
Nesvorn\'y, D., Vokrouhlick\'y, D., \& Morbidelli, A. 2007, \aj, 133, 1962

\bibitem[P\'erez et al.(2018)]{per18}P\'erez, L. M., Benisty, M., Andrews, S. N., et al. 2018, DSHARP X, \apjl, 869, L50

\bibitem[Pineda et al.(2018)]{pin18}Pineda, J. E., Szul\'agyi, J., Quanz, S. P., van Dishoeck, E. F., et al. 2018, \apj, arXiv:1811.10365

\bibitem[Reyes-Ruiz et al.(2015) ]{rey15}Reyes-Ruiz, M., Aceves, H., \& Chavez, C. E. 2015, \apj, 804, 91

\bibitem[Robutel \& Gabern(2006)]{rob06}
Robutel, P., \& Gabern, F. 2006, \mnras, 372, 1463

\bibitem[Roig \& Nesvorn\'y(2015)]{roi15}
Roig, F., \& Nesvorn\'y, D. 2015, \aj, 150, 186

\bibitem[Ruane(2017)]{rua17}Ruane, G., Mawet, D., Kastner, J., et al. 2017, \aj, 154, 73 

\bibitem[Tsiganis et al.(2005)]{tsi05}Tsiganis, K., Gomes, R., Morbidelli, A., \& Levison, H. F. 2005, Nature, 435, 459

\bibitem[van der Marel et al.(2019)]{vdm19}van der Marel, N., Dong, R., di Francesco, J., et al. 2019, \apj, arXiv:1901.03680

\bibitem[Vokrouhlick\'y et al.(2016)]{vok16}Vokrouhlick\'y, D., Bottke, W. F., \& Nesvorn\'y, D. 2016, \aj, 152, 39

\bibitem[Zellner(2017)]{zel17}Zellner, N. E. B. 2017, Orig Life Evol. Biosph., 47, 261

\bibitem[Zhang et al.(2018)]{zha18}Zhang, S., Zhu, Z., Huang, J., et al. 2018, DSHARP VII, \apjl, 869, L47

\end{thebibliography}
\end{document}